# Unraveling the Ai2 Asta Scholarly Research Assistant Citation System

Enrique Orduña-Malea*[1] and Carlos Lopezosa[2]

[1] The iMetrics Lab. Department of Audiovisual Communication, Documentation and History of Art, Universitat Politècnica de València, Valencia (Spain)
[2] Departament de Biblioteconomia, Documentació i Comunicació Audiovisual, Universitat de Barcelona, Barcelona (Spain)
* Corresponding author: enorma@upv.es

**Abstract**
Despite the growing integration of Deep Research tools into academic workflows, empirical evidence on the operation, stability, and potential biases of their citation systems remains scarce. This study addresses this gap by evaluating the intensity, consistency, and bibliographic characteristics of references cited in the literature reports generated by *Ai2 Asta*, with the aim of understanding how its citation system operates and assessing its implications for scholarly communication. To this end, ten domain-specific queries were submitted to Asta's Summarise Literature feature, and two independent rounds of data collection were conducted. From each report, in-text citations, cited references, as well as other metrics related to the response process were extracted and examined. The results reveal high citation intensity, with reports integrating numerous in-text citations grounded in retrieved evidence and a diverse yet concentrated set of venues. However, notable instability is observed in the composition of cited references across identical queries, alongside a lack of concordance between retrieved documents and those ultimately cited, suggesting additional opaque selection mechanisms during report generation. These findings indicate that, while *Ai2 Asta* produces well-structured and quality reports, its instability and opacity in the citation process, pose challenges in quantitative science studies due to their lack of reproducibility and transparency. Despite the restricted number of queries and disciplinary scope, the results offer valuable insights for researchers, bibliometricians, developers, and research evaluators seeking to understand, use or regulate AI-based scholarly assistants responsibly.

**Resumen:**
A pesar de la creciente integración de las herramientas de *Deep Research* en los flujos de trabajo académicos, siguen siendo escasas las evidencias empíricas acerca del funcionamiento, estabilidad y posibles sesgos de sus sistemas de citación. Este estudio aborda este hueco en la literatura evaluando la intensidad, consistencia y características bibliográficas de las referencias citadas en los informes de literatura generados por *Ai2 Asta*, con el objetivo de comprender cómo funciona su sistema de citación y analizar sus implicaciones para la comunicación científica. Para ello, se enviaron diez consultas específicas a través de la funcionalidad Summarise Literature de Asta, y se llevaron a cabo dos rondas independientes de recogida de datos. De cada informe se extrajeron y examinaron las citas embebidas en el texto, las referencias citadas, así como otras métricas relacionadas con el proceso de respuesta. Los resultados revelan una alta intensidad de citación, así como un conjunto de fuentes diverso pero concentrado en unas pocas fuentes. Sin embargo, se observa una notable inestabilidad en la composición de las referencias citadas en consultas idénticas, junto con una falta de concordancia entre los documentos recuperados y los finalmente citados, lo que sugiere la existencia de mecanismos adicionales y opacos de selección durante la generación del informe. Estos hallazgos indican que, aunque *Ai2 Asta* produce informes bien estructurados y de calidad, su inestabilidad y falta de transparencia en el proceso de citación, plantean desafíos de cara a su uso en estudios cuantitativos de la ciencia, debido a su falta de reproducibilidad y trazabilidad. A pesar del número limitado de consultas y del alcance disciplinar restringido, los resultados ofrecen información valiosa para investigadores, expertos en Bibliometría, desarrolladores y evaluadores científicos interesados en comprender, utilizar o regular de forma responsable los asistentes académicos basados en IA.

## Supplementary material

Available at https://riunet.upv.es/handle/10251/230713

## Statements and Declarations

*Funding:* Grant PID2022-142569NA-I00, funded by MCIN/AEI/10.13039/501100011033 and by "ERDF A way of making Europe"

*Author contributions:* EO: Conceptualization, Methodology, Formal analysis, Visualization, Writing-original draft ; CL: Conceptualization, Methodology, Writing-review & edit.

*Competing interests:* The authors have no competing interests to declare relevant to this article's content.

## Preprint

This manuscript is a preprint of a study accepted for publication in Revista Panamericana de Comunicación. Please cite as:

# 1 Introduction

The use of Large Language Models (LLMs) has rapidly gained popularity since the launch of ChatGPT, a conversational chatbot by OpenAI in November 2022, which significantly contributed to democratizing access to this type of technology for the public. ChatGPT reached 1 million users in its first five days and 100 million in its first two months of operation (Hu, 2023). Currently, it has reached 800 million weekly active users (Sor, 2025), and its website receives approximately 5.9 billion monthly visits, according to Similarweb data.[1]

LLMs have become integrated into the scientific workflow (Binz et al., 2025), reshaping and transforming research activity across all disciplines (Rane et al., 2023), including the social sciences (Grossmann et al., 2023). LLMs are being used in tasks such as writing support (Chen, 2023), literature synthesis (Zheng et al., 2023; Scherbakov, 2025; Silva and Wickramaarachchi, 2025), data analysis (Nejjar et al., 2025), assistance in designing research techniques such as interviews or surveys (Jansen et al., 2023), and peer review (Kousha and Thelwall, 2024; Sun, 2025).

In the specific case of searching publications or conducting literature reviews (systematic, scoping or narrative), the use of LLMs presents a series of limitations due to the characteristics of the data with which the LLM has been trained (e.g., scarcity of data, biases and lack of diversity), as well as its work processes (e.g., lack of transparency and instability in the results) (Rossi et al., 2024), including the generation of hallucinations and ghost references (Orduña-Malea and Cabezas-Clavijo, 2023; Walters and Wilder, 2023), the breakdown of the attribution chain (Codina, 2025), which can lead to obtaining false results, compromising the advancement of science. This issue is even enhanced due to the potential user's lack of literacy in the use of AI tools (Ng et al.,2021).

For this reason, AI tools designed to answer informational questions incorporate retrieval augmented generation (RAG), a technique that allows for the retrieval of relevant information from specific external knowledge bases in real time before answering user queries through LLM (Lewis et al., 2020). This is achieved through vectorized search, which involves breaking down both the query and the documents indexed in the knowledge base into a series of numerical values, called embeddings, represented by vectors, and then searching for a match between them using similarity indicators (e.g., Xian et al., 2024). This reduces inaccuracies and yields more precise, up-to-date, and transparent answers (Gao et al., 2024).

Although tools provided by renowned database platforms like Scopus AI or Web of Science AI Research Assistant utilize RAGs to generate their answers, RAG-based information retrieval systems often suffer from superficial document retrieval, lacking in-depth reasoning and source verification. However, advanced autonomous systems called Deep Research (DR) represent an evolution, integrating, on the one hand, a deep search based on iterative searches, combining vector search with other techniques to acquire, aggregate, and analyze external information, and, on the other hand, deep and dynamic reasoning and adaptive planning (Huang et al., 2025).

[1] https://www.similarweb.com/es/website/chatgpt.com

Despite producing slower searches, DR systems have demonstrated the capacity to process large amounts of information and generate more accurate reports (Xu and Peng, 2025). DR tools are increasingly used by researchers to conduct comprehensive literature reviews and synthesize existing knowledge (Xu et al., 2025).

Since 2023, a large number of DR tools, both commercial and non-commercial, have been launched (Tay, 2025). Among these, we can distinguish those of general purpose, but which can be applied to a scientific context (e.g., *ChatGPT Deep Research*, *Gemini Deep Research* or *Perplexity Labs*), and those specifically oriented to the scientific community, among which stand out *Elicit*, *Scispace Deep Review*, *Future House*, *Consensus Deep Research*, *Undermind*, *AnswerThis*, *PagerDigest*, *Scopus Deep Research* (available as part of *Scopus AI*), and recently *Ai2 Asta* (aka *Scholar QS*).

Thanks to the combination of Deep Search and RAG, Deep Research tools not only enhance transparency in the process of generating the final report by explicitly mentioning the publications selected during the search process, but also facilitate recognition and intellectual attribution to the authors of these publications.

Arguably, the increasing use of DR tools by researchers to generate brief literature reviews may lead to publications included in the RAG report, generated in response to a specific question, being more likely to be cited in the publications of the tool's users. In other words, the Deep Search + RAG process may affect not only the search processes for scientific literature but also the literature ultimately cited by researchers, and therefore, the citation-based impact of publications, venues, and authors.

To study the potential consequences of using deep research tools on scientific impact, it is necessary to determine the coverage of sources used by these tools, as well as the possible existence of biases in the selection of specific publications and sources for particular queries. The primary objective of this exploratory work is to determine the quantity, variety, and bibliographic characteristics of the publications utilized by a deep research tool, using the *Ai2 Asta* tool as a case study. Specifically, the following research questions are drawn:

**RQ1.** What is the citation intensity? In other words, how many bibliographic references does a report include?

**RQ2.** What is the stability of the citation system? In other words, what is the variability in the publications cited in a report for the same question asked multiple times?

**RQ3.** What is the citation diversity? In other words, what are the bibliographic characteristics of the cited publications (year, venue, citation-based impact)?

The remainder of the work is structured as follows. Chapter 2 provides a basic description of the operation and features of *Ai2 Asta*. Chapter 3 describes the methodological process followed to obtain the references mentioned in the tool's reports, as well as the collection of bibliographic data from these publications.

Chapter 4 presents the main results obtained, which are discussed in Chapter 5. Finally, Chapter 6 presents the concluding remarks.

## 2 The Deep Research tool: Ai2 Asta

*Ai2 Asta* is an artificial intelligence ecosystem launched on August 25, 2025, by The Allen Institute for Artificial Intelligence (Ai2), geared towards use in scientific research,[2] and composed of three fundamental components: *Asta*[3] (a set of discovery, synthesis, and analysis tools designed to support researchers' needs), *AstaBench*[4] (a benchmarking framework for evaluating and comparing AI agents), and *Asta Resources*[5] (a set of tools, baseline agents, templates, and APIs for developers to build, test, and refine scientific AI agents). Asta agents[6] and the evaluation framework[7] are open-source.

Specifically, *Ai2 Asta* integrates two functionalities (*Find Papers* and *Summarize Literature*), which come from two previously separate tools: *AI2 Paper Finder*[8] (Deep Search) and *Scholar QS*[9] (Deep Research). A third functionality (*Analyze Data*), currently available in beta for select partners, will be added. These functionalities together form a scholarly research assistant with broad and deep coverage across 23 fields of study, updated weekly.

The system starts by understanding and re-formulating the user's query to created two queries to be submitted to the *Semantic Scholar* API (Kinney et al., 2023). The first, it is a keyword-based query submitted to the keyword search over paper abstracts (around 108 million abstracts) endpoint. The second is a semantic-based query, which is submitted to the text snippets search endpoint, which contains snippets (brief pieces of text of around 500 terms) from full-text open-access papers drawn from a paper's title, abstract, and body text (over 12 million full-text papers). After an internal process applying ranking algorithms (weighted sum of embedding similarity and bm25 scores) and neural re-ranking (Singh et al., 2025), the system selects the 50 most relevant passages to answer the query from up to 256 snippets and 20 abstracts. Each passage is then analyzed to extract the most relevant and precise quotes. These quotes are thematically classified into different sections, which are used to structure the response. Finally, claims in the answer are supported by in-text citations, which can be clicked to reveal the cited papers title and authors, allowing for quick verification of the claim (Singh et al., 2025). Cited sources are either the quotes assigned to the section or abstracts of papers that are cited within these quotes.

Additionally, the Ai2 Lab team has launched *Asta Summary Citation Counts*,[10] an open dataset that compiles the most frequently cited publications in reports generated by the *Summarize Literature* tool. This dataset is based on an analysis of 113,000 queries, which collect approximately 4 million citations from around 2 million publications and are updated weekly. To our knowledge, this dataset is unique in its category.

[2] https://allenai.org/blog/asta
[3] https://asta.allen.ai
[4] https://allenai.org/asta/bench
[5] https://allenai.org/asta/resources
[6] https://github.com/allenai/agent-baselines
[7] https://github.com/allenai/asta-bench
[8] https://allenai.org/blog/paper-finder
[9] https://allenai.org/blog/ai2-scholarqa
[10] https://huggingface.co/datasets/allenai/asta-summary-citation-counts

*Asta Summary Citation Counts* allows us to understand Asta's citation patterns, which is very helpful to the quantitative science studies community in understanding how Deep Research tools identify and use scientific works to support answers to user questions. However, while this data is useful for understanding citation patterns at the tool level, it does not enable us to understand citation patterns at the query level. In other words, it does not allow us to analyze publications cited in reports linked to queries aimed at synthesizing the literature on a specific topic, and whose presence in the Deep Research report could facilitate the citation of that work. For this reason, it is necessary to conduct specific queries and then to analyze the cited publications—precisely the focus of this study.

## 3. Methods

### 3.1. Data collection

To answer the research questions, 10 specific queries were designed to be run against Asta's "Summarize literature" function, in order to verify the references cited in the generated reports (Table 1). The questions relate to topics of broad debate within the field of quantitative science studies and, in some cases, have generated controversy and opposing viewpoints, which may facilitate the creation of comparative tables and interpretive reports.

**Table 1**
List of literature review questions

| N | Question |
|---|---|
| **1** | What is the current state of the Webometrics (also known as Cybermetrics) discipline in 2025, considering its advantages, limitations, current applications, and main lines of research? |
| **2** | How effective is the h-index in assessing researchers' impact, and what are its main advantages, limitations, and existing variants? |
| **3** | How can the societal impact of research be measured through Altmetrics, and what are the main advantages and disadvantages of this approach? |
| **4** | How Bibliometrics has contributed to understanding the conceptual structure and evolution of citizen science research? |
| **5** | What are the strengths and limitations of using major bibliographic databases for bibliometric analyses, particularly regarding document and citation coverage, metadata quality, and search functionalities? |
| **6** | What are the effects of open access publishing on the citation impact of scientific publications, and how robust is the evidence supporting the open access citation advantage? |
| **7** | What insights have bibliometric studies provided on gender disparities in science, and what methodological approaches have been most effective in addressing this issue? |
| **8** | How have bibliometric studies explored the mobility of researchers, and what are the key findings, indicators, and methodological challenges? |
| **9** | How are university rankings constructed based on bibliometric indicators, and what are |
| | the main uses, quality concerns, biases, and advantages or disadvantages of these rankings? |
| **10** | What are the main schools of thought in the evaluation of scientific activity, and what are the advantages and disadvantages associated with each evaluation approach? |

To respond to RQ1 and RQ2, the queries were launched directly into the Asta text box, selecting the "summarize literature" option.[11] For each question, the search process information provided by *Ai2 Asta* (used to extract search duration, number of retrieved documents and field of study), the generated report in JSON format (used to extract the number of created sections and comparative tables, and the number of embedded citations), and the cited references report in BIB format (used to extract bibliographic information from each reference, specifically publication name, publication year, citations received, title, authors, and publication ID) were collected.

To address RQ3, two data collections were conducted. The first was performed on October 29th and the second on November 2nd, using the same configuration, machine and location in order to minimize external variables due to technical issues.

### 3.2. Data analysis

The collected data were exported to a spreadsheet for basic descriptive statistical analysis. Citation discrepancies between data sets were visualized using Venn diagrams created with *Venn Diagram Plotter*,[12] and data figures were generated using *Scimago Graphica* (Hassan-Montero et al., 2022).[13]

Additionally, general contextual information was extracted from the *Asta Summary Citation Counts*. The file "sqa_citation_ranking_all_time.parquet" was downloaded on October 28th. General data on venues and publication years for the publications cited in Asta were extracted using an ad hoc Python script.

## 4. Results

### 4.1. Citation intensity

Table 2 presents the raw data obtained from the two rounds of data collection. As shown, the mean number of cited references per report in the first data collection is 22.2, although the dispersion is considerable (ranging from 8 citations in Q1 to 42 in Q2). The results from the second data collection are highly similar and strongly correlated ($R_p = 0.86$), although in this case the mean number of cited references per report is slightly higher ($\tilde{x} = 27.3$).

[11] https://asta.allen.ai/discover
[12] https://pnnl-comp-mass-spec.github.io/Venn-Diagram-Plotter
[13] https://www.graphica.app/

**Table 2**

Search parameters collected from *Ai2 Asta* reports generated as a response to academic questions

**DATA COLLECTION 1**

| Query | Field of study | Time | Passages | Abstracts | Relevant papers | Papers cited | In-text citations | Extension (words) | Sections | Comparative Tables | Bullet sections |
|---|---|---|---|---|---|---|---|---|---|---|---|
| **1** | Computer Science | 1m 49s | 250 | 19 | 25 | 8 | 38 | 1461 | 6 | 1 | 1 |
| **2** | Sociology, Education | 3m 8s | 248 | 5 | 30 | 40 | 94 | 1931 | 5 | 1 | 1 |
| **3** | Sociology, Business | 2m 56s | 250 | 14 | 25 | 24 | 52 | 2406 | 5 | 2 | 1 |
| **4** | Sociology, Education | 3m 1s | 252 | 15 | 24 | 17 | 59 | 1864 | 6 | 1 | 1 |
| **5** | Computer Science | 3m 2s | 255 | 0 | 42 | 27 | 69 | 2467 | 5 | 0 | 0 |
| **6** | Sociology, Education | 2m 42s | 249 | 8 | 35 | 38 | 81 | 2488 | 6 | 0 | 0 |
| **7** | Sociology, Education | 4m 1s | 244 | 15 | 35 | 36 | 91 | 3129 | 8 | 0 | 1 |
| **8** | Sociology, Business | 2m 57s | 254 | 17 | 21 | 24 | 94 | 3002 | 5 | 0 | 0 |
| **9** | Education, Business | 4m 12s | 255 | 0 | 23 | 26 | 109 | 3090 | 7 | 1 | 0 |
| **10** | Education, Sociology | 2m 25s | 235 | 17 | 45 | 22 | 56 | 1792 | 5 | 0 | 0 |

**DATA COLLECTION 2**

| Query | Field of study | Time | Passages | Abstracts | Relevant papers | Papers cited | In-text citations | Extension (words) | Sections | Comparative Tables | Bullet sections |
|---|---|---|---|---|---|---|---|---|---|---|---|
| **1** | Computer Science, Sociology | 2m6s | 252 | 19 | 24 | 9 | 35 | 1500 | 6 | 1 | 2 |
| **2** | Sociology, Education | 2m39s | 250 | 6 | 30 | 42 | 113 | 2382 | 5 | 0 | 0 |
| **3** | Sociology, Business | 2m 57s | 250 | 14 | 25 | 35 | 74 | 2137 | 5 | 1 | 2 |
| **4** | Sociology, Education | 1m 46s | 252 | 15 | 23 | 15 | 48 | 1770 | 5 | 0 | 0 |
| **5** | Computer Science, Linguistic | 2m 43s | 255 | 4 | 42 | 32 | 79 | 3071 | 6 | 0 | 0 |
| **6** | Sociology | 2m 31s | 252 | 6 | 26 | 29 | 58 | 2373 | 6 | 0 | 0 |
| **7** | Sociology, Education | 2m 34s | 243 | 16 | 33 | 38 | 70 | 2407 | 5 | 0 | 0 |
| **8** | Sociology, Education | 2m 40s | 252 | 18 | 22 | 24 | 85 | 2565 | 5 | 1 | 1 |
| **9** | Education, Business | 2m 44s | 253 | 0 | 25 | 23 | 88 | 2463 | 5 | 0 | 0 |
| **10** | Education, Sociology | 3m 0s | 235 | 18 | 47 | 26 | 87 | 3716 | 7 | 0 | 0 |

Note: extension only considers the “sections- text” tag of the JSON file with the report generated.

The reports are lengthy, ranging from 1,500 to 4,000 words (excluding abstracts and section titles), with longer reports generally — though not always — containing a greater number of cited references. The number of in-text citations (or claims) per report is comparatively high ($\tilde{x}$ = 78.3 in the first data collection; $\tilde{x}$ = 73.7 in the second) in relation to the number of cited references. This indicates a high rate of citation per reference, meaning that publications tend to be cited repeatedly throughout the report. Only a small number of errors (i.e., unresolved links to cited publications) were detected in the generation of in-text citations. In addition, the presence of claims not supported by retrieved evidence—labelled as "LLM memory" when generated in the absence of relevant passages—was noted, particularly in the first data collection, and was virtually absent in the second.

A large number of passages was retrieved in both data collections ($\tilde{x}$ = 249), very close to the system maximum (256). In general, queries with longer search durations tended to generate a higher number of in-text citations and, indirectly, a higher number of cited references. However, this pattern was not universal. Otherwise, abstracts were used less frequently, especially in cases where fewer passages were retrieved and fewer in-text citations appeared in the final reports.

No relationship was found between the number of relevant publications retrieved during the deep search process and the number of references ultimately cited in the reports. This suggests that publications continue to be added or discarded during the final report generation stage. Indeed, in 50% of the queries, the number of cited references exceeded the number of relevant publications retrieved during deep search.

### 4.2. Citation Stability

The total number of cited references differs between the first and second data collections, depending on the query. Moreover, the variation is not limited to the number of references but also affects their composition. For instance, of the eight references cited in the report generated for Query #1 in the first data collection, only four are also present among the nine references cited for the same query in the second data collection.

The degree of overlap varies substantially across queries, ranging from 34.2% in Query #6 (where only 13 of the 38 references cited in the first data collection appear in the second) to 95.5% in Query #10 (where 21 of the 22 references cited in the first data collection are also included in the second). A full overview of the overlap in cited references across all queries is presented in Figure 1.

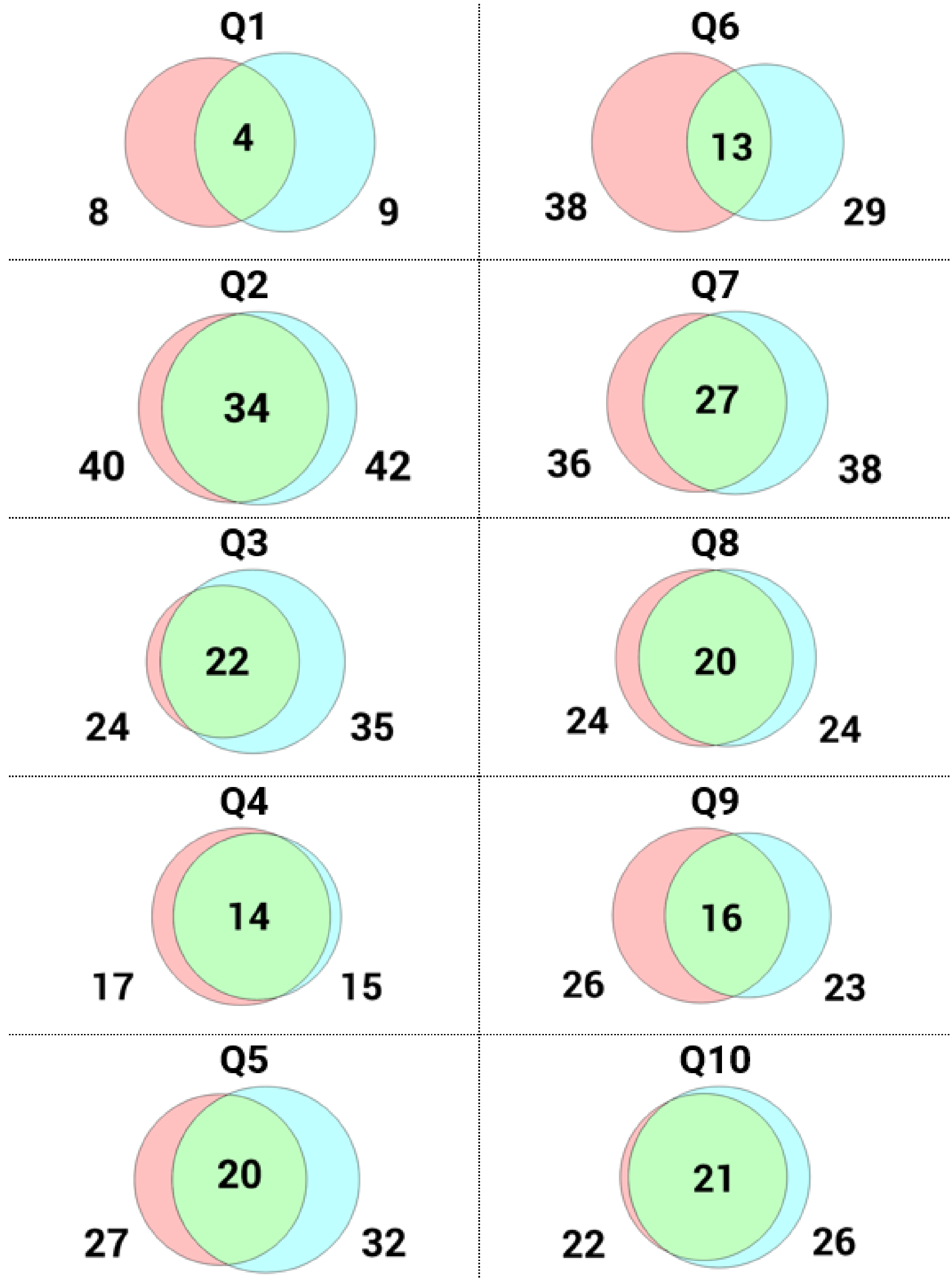

**Figure 1**
Venn Diagram of cited references in *Ai2 Asta*

Note: the red circle corresponds to the number of cited references collected in the first sample; the blue circle corresponds to the number of cited references collected in the second sample; the green circle corresponds to share cited references in both samples.

In four cases, the assigned field of study changed slightly between the two data collections (e.g., Query #8 shifted from "Sociology/Business" to "Sociology/Education"). Such changes may alter the deep search process and lead to the retrieval of different references. However, discrepancies were also observed in queries where the field of study remained identical across both data collections.

### 4.3. Citation Diversity

The ten queries generated a total of 262 cited references in the first data collection, drawn from 105 unique sources, with a small number of journals accounting for most citations (only 19% of sources were cited more than once). The figures are similar in the second data collection, which resulted in 273 cited references from 111 unique sources, 20.7% of which were cited more than once.

The venues providing the highest number of cited references were Scientometrics, PLoS One, and arXiv.org (Table 3). These three sources collectively account for a substantial proportion of cited references in both data collections (29% and 30.8%, respectively). Other frequently cited venues include JASIST (specialised in information science, including but not limited to bibliometrics), PNAS and IEEE Access (both multidisciplinary), Journal of Informetrics and Quantitative Science Studies (bibliometrics-focused), and Publications (specialised in scholarly publishing more broadly).

**Table 3**
Venues most frequently cited in *Ai2 Asta* reports for the selected queries

| **Data collection 1** | | **Data collection 2** | |
|---|---|---|---|
| **Venue** | **N** | **Venue** | **N** |
| Scientometrics | 41 | Scientometrics | 44 |
| PLoS ONE | 20 | PLoS ONE | 20 |
| arXiv.org | 15 | arXiv.org | 20 |
| JASIST | 13 | JASIST | 11 |
| PNAS | 9 | PNAS | 10 |
| Journal of Informetrics | 8 | Journal of Informetrics | 9 |
| Quantitative Science Studies | 7 | Quantitative Science Studies | 6 |
| Publications | 7 | IEEE Access | 6 |
| IEEE Access | 5 | Publications | 5 |
| Cureus | 4 | Frontiers in Research Metrics and Analytics | 5 |

The number of venues cited varies considerably between queries (Table 4), ranging from a minimum of eight (Query #1 in the first data collection) to a maximum of 27 (Query #2 in the second).

**Table 4**
Venues and citation-based impact in *Ai2 Asta* reports' cited references

| | DATA COLLECTION 1 | | | | | DATA COLLECTION 2 | | | | |
|---|---|---|---|---|---|---|---|---|---|---|
| QUERY | Cited references | Unique venues | Recent publications | Avg. citations | Median citations | Cited references | Unique venues | Recent publications | Avg. citations | Median citations |
| #1 | 8 | 8 | 1 (12.5%) | 11 | 4 | 9 | 9 | 4 (44.4%) | 15.6 | 4 |
| #2 | 40 | 26 | 5 (12,5%) | 286.5 | 21 | 42 | 27 | 4 (9.5%) | 288.4 | 21 |
| #3 | 24 | 15 | 4 (16.7%) | 81.2 | 15 | 35 | 21 | 4 (11.4%) | 117.3 | 15 |
| #4 | 17 | 12 | 4 (23.5%) | 447.9 | 37 | 15 | 12 | 4 (26.7%) | 512.5 | 37 |
| #5 | 27 | 18 | 9 (33.3%) | 334.8 | 24 | 32 | 21 | 13 (40.6%) | 178.9 | 24 |
| #6 | 38 | 20 | 7 (18.4%) | 150.9 | 62 | 29 | 13 | 5 (17.2%) | 235.9 | 62 |
| #7 | 36 | 20 | 15 (41.7%) | 115.4 | 22 | 38 | 20 | 15 (39.5%) | 102.1 | 22 |
| #8 | 24 | 13 | 5 (20.8%) | 24.1 | 13.5 | 24 | 15 | 5 (20.8%) | 20.4 | 13.5 |
| #9 | 26 | 17 | 9 (34.6%) | 48.8 | 7.5 | 23 | 19 | 6 (26.1%) | 42.8 | 7.5 |
| #10 | 22 | 11 | 1 (4.5%) | 37.4 | 18 | 26 | 14 | 1 (3.8%) | 49.7 | 18 |

Note: recent publications correspond to publications from 2023 onwards.

The proportion of recent publications (from 2023 onwards) among the cited references is noteworthy, although it fluctuates across queries and between data collections. For instance, in Query #1 it increases from 12.5% in the first data collection to 44.4% in the second, whereas in Query #7 it remains consistently high in both (41.7% and 39.5%, respectively). Further detail on the presence of recent publications is provided in Figure 2.

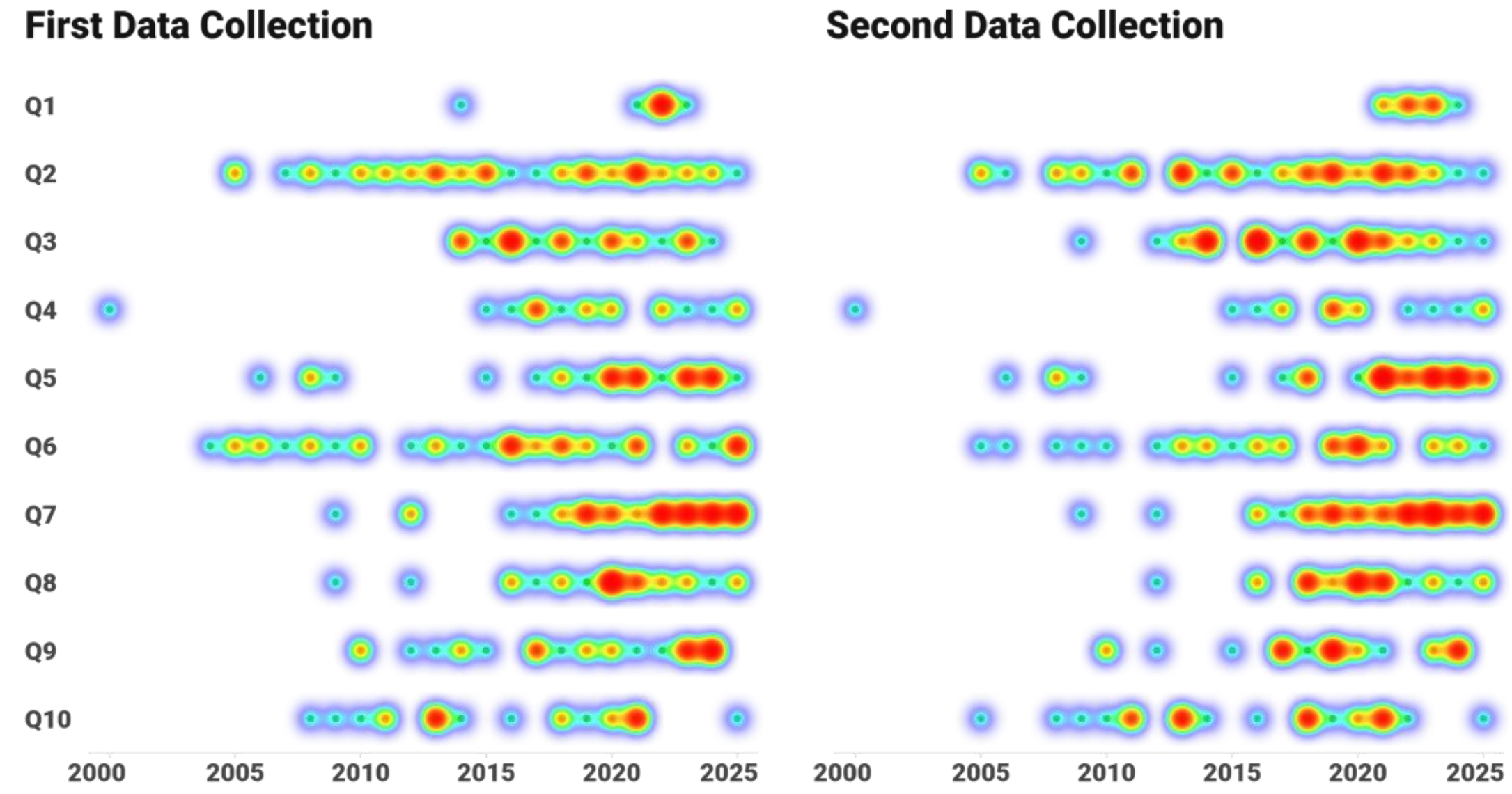

**Figure 2**

Distribution of publications cited in *Ai2 Asta* reports per query according to the year of publication

In addition to the high proportion of recent publications (which tend to have lower accumulated citation counts), the reports also include a substantial number of highly cited works. In the first data collection, 56 of the 262 cited references (21.4%) had received more than 100 citations, while 28 had received none. Comparable figures were observed in the second data collection, with 57 references exceeding 100 citations and 29 receiving zero. These results account for the marked differences between the mean and median citation values across queries, as shown in Table 3. The citation distribution per query is displayed in Figure 3.

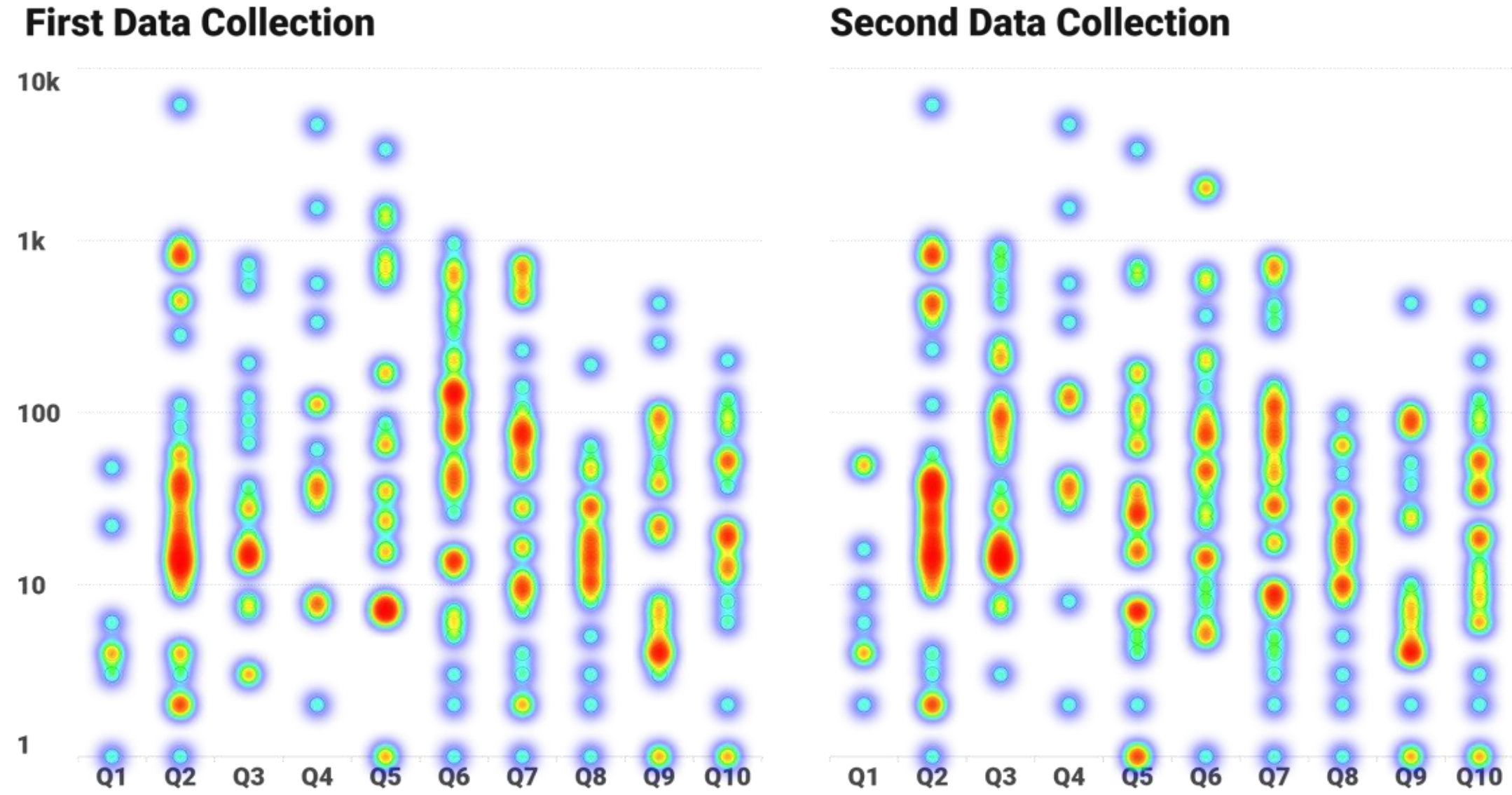

**Figure 3**
Distribution of publications cited in *Ai2 Asta* reports per query according to the number of citations received

## 5. Discussion

The citation system used by *Ai2 Asta* has been analysed through ten specific queries oriented to the field of science studies, in order to ascertain the intensity and diversity of the citations embedded in the reports generated, as well as the variability of the system, constituting the first study to date in which the citation system of a deep research tool is analysed from an informetric point of view.

Regarding the citation intensity (RQ1), the reports tend to include approximately between 20 and 40 cited references, which is considered a high number given the length of the reports (around 2.3k-2.4k words on average). Some of these cited references are repeated quite frequently in the text, resulting in a large number of in-text citations. Consequently, a few works may have a significant relevance on the report's content. Otherwise, prompts with fewer cited references could be a consequence of the question's formulation rather than simply due to less coverage in Semantic Scholar. For example, Query #1 has a time constraint (the year 2025) that could explain the low number of cited references found, making the clarity, precision, and conciseness of the prompt essential for evaluating the results. In any case, the purpose of this functionality is to provide the user with a generic report that answers the question posed in a well-supported manner, rather than offering a systematic and exhaustive review of the topic. In this respect, the "Find Paper" functionality (not analyzed in this study) is more comprehensive in locating scientific literature.

Formulating the same query at two different times has allowed us to verify the existence of significant variability in the references cited in the generated reports (RQ2). This approach differs from traditional bibliographic databases and academic search engines, as the system does not transparently display all relevant publications found. Instead, it only includes those ultimately selected to generate the report, and these selections vary significantly depending on when the query is performed. This procedure creates uncertainty within the system, since the same query can yield different results, both in terms of content and cited references.

Regarding the references cited in the reports (RQ3), the results show a significant concentration of journals. While the appearance of journals such as *Scientometrics*, *JASIST*, *Journal of Informetrics*, or *QSS* among the most cited sources is logical, given the nature of the questions, the results show an excessive presence of multidisciplinary journals (*PLoS One*, *PNAS*, *IEEE Access*) and the *arXiv.org* repository. However, certain limitations in source identification are observed, which restrict the accuracy of raw venue-related statistics. For example, some cited references lack accurate metadata. For example, 23 cited references in the first data collection and 19 in the second do not provide source information. Otherwise, some normalization problems are observed (e.g., "Proceedings of the National Academy of Sciences of the United States of America" and "Proc. Natl. Acad. Sci. USA"; and "J. Assoc. Inf. Sci. Technol." and "Journal of the Association for Information Science and Technology").

*Ai2 Asta* performs the deep search process through *Semantic Scholar*. Therefore, the shortcomings in the coverage of this database constitute Asta's first inherent bias, as well as the biases of *Semantic Scholar Academic Graph*, which "emphasizes direct title matches and highly-cited papers with recent publication dates" (Kinney et al., 2023). Furthermore, the system positively favors researchers in fields with a greater number of publications available on *arXiv*[14].

To verify the tool's inherent biases, the raw *Ai2 Asta* data (*Asta Summary Citation Counts*) were analyzed to identify the most frequently included publications in the generated reports (Figure 4), the most frequently cited venues (Table 5), and the most frequent assigned fields of study across all queries received by the tool (Table 6). The analysis reveals a high use of works published on arXiv.org and *PLoS One* in report generation, with a preponderance of topics in medicine and computer science. Furthermore, the data indicate that *Ai2 Asta* has cited 2,100,007 publications, of which 55.2% (1,158,500) were published since 2020. Therefore, these biases are inherited in the specific queries performed, an aspect that must be taken into account.

[14] https://allenai.org/blog/ai2-scholarqa

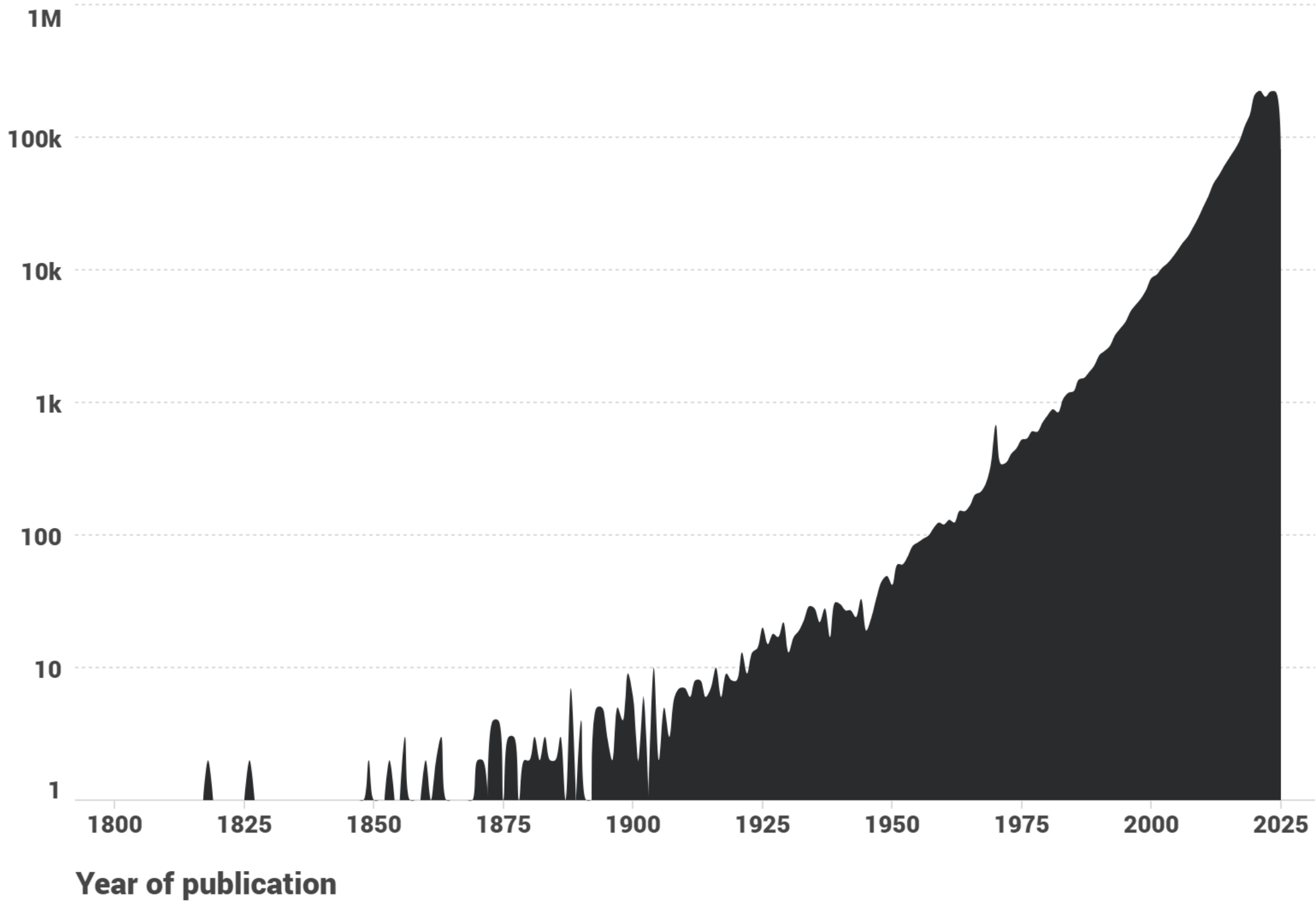

**Figure 4**
Number of publications cited in *Ai2 Asta* according to the year of publication

**Table 5**
Ranking of Journals in *Ai2 Asta*

| Journal | Publications | Queries | Citations |
|---|---|---|---|
| arXiv.org | 53,903 | 118,943 | 160,419 |
| PLoS ONE | 32,345 | 50,752 | 72,910 |
| Scientific Reports | 20,392 | 30,805 | 44,001 |
| Sustainability | 16,485 | 37,132 | 53,875 |
| Frontiers in Psychology | 15,586 | 36,080 | 50,695 |
| International Journal of Molecular Sciences | 15,088 | 25,586 | 37,851 |
| International Journal of Environmental Research and Public Health | 14,431 | 27,789 | 40,672 |
| IEEE Access | 11,778 | 22,822 | 32,734 |
| Italian National Conference on Sensors | 11,716 | 20,225 | 29,751 |
| bioRxiv | 11,006 | 15,228 | 20,920 |
| Molecules | 8,931 | 15,717 | 23,368 |
| Cureus | 7,701 | 12,812 | 19,380 |
| Materials | 7,356 | 12,139 | 18,041 |
| Applied Sciences | 7,224 | 13,054 | 19,153 |
| Heliyon | 7,173 | 14,632 | 21,385 |
| Journal of Physics: Conference Series | 6,953 | 10,059 | 14,051 |
| Nutrients | 6,926 | 12,270 | 18,422 |
| Nature Communications | 6,463 | 9,823 | 13,759 |
| IOP Conference Series: Earth and | 6,300 | 10,115 | 14,771 |
| Environment | | | |
| Journal of Biological Chemistry | 6,230 | 8,091 | 10,916 |
| E3S Web of Conferences | 5,733 | 10,320 | 14,689 |
| PNAS | 5,555 | 8,684 | 12,119 |
| Frontiers in Immunology | 5,535 | 8,861 | 12,936 |
| Energies | 5,393 | 9,234 | 13,460 |
| Annual Meeting of the Association for Computational Linguistics | 5,305 | 14,671 | 19,200 |
| Journal of Clinical Medicine | 5,134 | 8,494 | 12,720 |

**Table 6**
Ranking of Fields of Study in *Ai2 Asta*

| Fields Of Study | Number of queries |
|---|---|
| Medicine | 300,908 |
| Computer Science | 112,680 |
| Biology, Medicine | 94,273 |
| Psychology | 65,552 |
| Environmental Science | 52,084 |
| Biology, Environmental Science | 47,030 |
| Engineering, Environmental Science | 45,361 |
| Medicine, Environmental Science | 41,670 |
| Psychology, Medicine | 39,362 |
| Education | 36,961 |
| Agricultural and Food Sciences | 36,378 |
| Computer Science, Engineering | 35,285 |
| Medicine, Computer Science | 32,867 |
| Economics, Business | 28,986 |
| Computer Science, Education | 28,219 |
| Sociology | 28,114 |
| Biology, Medicine, Environmental Science | 26,560 |
| Engineering, Materials Science | 25,747 |
| Medicine, Engineering | 25,258 |
| Environmental Science, Agricultural and Food Sciences | 24,929 |
| Biology | 22,847 |
| Business | 22,290 |
| Psychology, Education | 21,785 |
| Computer Science, Business | 20,414 |

## 6. Conclusions and future research

The findings of this study indicate that *Ai2 Asta*'s citation system displays a distinctive combination of high citation intensity, moderate bibliographic diversity, and considerable instability across repeated queries. Although the tool produces well-structured reports enriched with numerous citation-backed claims—and can therefore serve as a valuable aid for researchers when used critically and appropriately—the underlying set of cited publications varies substantially even when identical queries are issued at different times. Moreover, the disconnect between the documents retrieved during deep search and those ultimately cited suggests the presence of additional, opaque selection mechanisms during report generation. These results carry important implications for scholarly practice, raising concerns about the reproducibility and traceability of AI-generated literature syntheses, highlighting the risk that unstable citation patterns may inadvertently shape researchers' citation behaviour, and underscoring the need for greater transparency in AI-driven retrieval and attribution processes. Overall, while *Ai2 Asta* offers high-quality and informative reports that can significantly support early-stage literature exploration, its limitations call for a cautious and informed use, as well as continued research into its technical behaviour and epistemic impact.

This research represents a first approach to understanding how *Ai2 Asta*'s citation system operates, especially in controlled scenarios. However, given the complexity of the processes involved (from the moment we issue the instruction to the moment we obtain the results) it is essential to explore new avenues of research that broaden and deepen the findings obtained.

Many aspects still remain to be examined, both in terms of in-depth document retrieval and the final selection of references during report generation, in order to obtain a more complete picture of its behavior. Therefore, future research could aim to expand the number and variety of queries analysed, incorporating different scientific disciplines and research topics. This increase would help determine whether the variability patterns detected in our study persist in contexts with different publication dynamics or whether, on the contrary, specific behaviors emerge depending on the field of knowled.

In addition, it would be interesting to include new dimensions of analysis focused, for example, on the overall quality of the reports generated by *Ai2 Asta*, considering aspects such as (1) the length of the responses, (2) their structural coherence, (3) the strength of the argumentation, and (4) the possible presence of biases and/or inaccuracies. This would allow for a more precise evaluation of how the Allen Institute for AI's tool constructs its syntheses and the extent to which these are reliable from an academic standpoint.

On the other hand, a detailed examination of the suitability and reliability of the cited references, for example, assessing whether the reference is properly constructed and genuine, whether the source has clear authorship and scholarly rigor, and whether the citations are appropriate and relevant to the content returned by the tool, would make it possible to determine whether the instability observed in our study is due to technical factors or to more complex processes of information selection.

Finally, we consider it important to develop longitudinal studies that analyze Asta's behavior over time, since new versions of the system will likely be developed to improve its performance. This approach will facilitate, on the one hand, the identification of trends related to model updates, corpus expansions, and changes in retrieval mechanisms, and on the other, the assessment of their impact on reproducibility, transparency, and the responsible use of its Deep Research resource for accelerating research processes.

The results obtained in this work may be of particular help to experts in bibliometrics and science studies, by enabling a deeper understanding of how scholarly research assistants operate and construct their citation systems, as well as to researchers and evaluators, who may better identify the potential limitations, strengths, and characteristic features of the reports generated by such tools. Moreover, these findings can also inform developers of AI-based scholarly assistants, offering empirical evidence about the citation process that may guide improvements in retrieval mechanisms. By highlighting both the capabilities and shortcomings of *Ai2 Asta*'s current behaviour, this study provides actionable insights that can support the design of more reliable, robust, and academically responsible Deep Research systems.